\documentclass{article}

\usepackage[utf8]{inputenc}

\usepackage{amsmath}
\usepackage{xfrac}
\usepackage{mathrsfs}

\title{Quantum Measurement Theory as simply explained by \\ N G van Kampen }
\author{M G Burt\\
\\Department of Physics, Durham University, Durham, DH1 3LE, UK}
\date{7th May 2022}

\begin{document}
\maketitle
\begin{abstract}
A translation is provided of the nonspecialist article by N G van Kampen `Het Postmoderne Obscurantisme' [ Postmodern Obscurantism ] (Nederlandse Tijdschrift voor Natuurkunde [Dutch Journal of Physics] $\boldsymbol{56}$ (2) p13 (1990)) which seeks to dispel much of the mysticism surrounding quantum measurement. The translation offered is `free' using appropriate idiom and figurative speech in keeping with the original author's inimitable style and superb command of English as displayed in his numerous publications. Square brackets have been used to contain extra clarification added by the translator and sometimes just to provide an alternative, more literal, translation to the idiomatic English used so as to help non-native English speakers.The original paper is put into context immediately below, followed by the translation itself.
\end{abstract}
\begin{center}
    \textbf{Context} 
\end{center}
 In his famous article “Against Measurement" ( \textit{Physics World} August 1990 p33), John Bell quoted and criticised three sources to illustrate his concerns about the then current understandings of the foundation of quantum mechanics. These sources were two books, the one by Landau and Lishitz, the other that by Gottfried, and a paper by N G van Kampen ( reference 5 in the translation below). Van Kampen responded robustly and cogently to the criticism in two letters to \textit{Physics World} in October 1990 ( page 20 ) and December 1991 (page 16 ). He also summarised his position in a letter to the \textit{American Journal of Physics} (\textbf{76} (11) November 2008). While these letters are useful awareness raising exercises, they are hardly adequate to satisfy the inquiring reader. On the other hand, his detailed theoretical publications can hardly be described as accessible to most physicists. So there is a clear need for an exposition suitable for the general reader. Such an exposition, entitled `Het Postmoderne Obscurantisme' [ Postmodern Obscurantism ], was published by van Kampen in Dutch. It was written in a very engaging and practical tone and its translation into English, to benefit the world-wide community of physicists, is long overdue.\\
For the reader more interested in the detail, an essential part of van Kampen's understanding of the measurement process is based on the relation between the microscopic description ( based on probability amplitudes ) and the macroscopic description ( based on probabilities alone) of a macroscopic system. While this is the subject of reference 7, it should be mentioned that this topic has been treated by van Kampen more recently in  \textit{Physica} A $\boldsymbol{194}$ (1993) pp 542 - 550 and in his book \textit{Stochastic Processes in Physics and Chemistry} (third edition 2007 Elsevier)  pp 451 - 456 .
\\
\\
\begin{center}
    \textbf{\Huge{Postmodern Obscurantism}}
\end{center}

\begin{center}
    \textbf{\huge{by N G van Kampen}}
\end{center}

\emph{\\“In philosophy one must always proceed from puzzlement and wonder to  enlightenment, that is to say, one must study and research for as long as it takes, so  that eventually, that which initially appears strange to us no longer appears so; but in theology one must proceed from a state of no wonder to a state of wonder and awe, that is to say, one studies and researches the Scriptures for as long as it takes so that what initially appears as mundane is perceived as heavenly and miraculous”
\\
\\
Isaac Beeckman}
\section{Introduction}

Obscure describes those who are unable to express themselves clearly.  Obscurants denotes those who do not \emph{want} to express themselves clearly; those who, for what ever reason, present things in a murkier form than is necessary. Traditionally, that has been considered a hallmark [ characteristic, even a virtue,] of theologians, as one readily sees from a reading of the second part of the quotation above. But here we are talking about the first half (philosophy must here be understood as natural science) and the fact that this [the message in the first half of the quotation] is so often trampled underfoot. Often statements about physics are given a mystical aura, either to pull the wool over the reader's eyes [mislead the reader] so as to promote some religious or philosophical, but unscientific, opinion, or merely just to sound interesting. This phenomenon is not new, but, in recent years, there has been an explosion of popular literature in which physics has been presented as something mysterious. That is postmodern obscurantism.

Quantum mechanics is especially susceptible to this phenomenon due to slogan-like declarations such as the following. (i) \emph{During a measurement an observer intervenes in the natural course of events}; nature has no reality independent of the observer. (ii) \emph{The observer influences the observed system}; there is no objectivity. (iii) \emph{ The wave function does not describe the state of the system, but the knowledge the observer has about the system}; that is why the wave function jumps on measurement and becomes a completely new function consistent with the information acquired.

These slogan-like declarations are not entirely untrue, and in the right context can be useful to illustrate certain aspects of quantum mechanics. From the voluminous literature I can only refer to some reviews $^{[1]}$. But, to make an impression on philosophers and other outsiders, these declarations are often assigned an absolute truth, which causes much confusion. On the basis of such misunderstood declarations, quantum mechanics has incurred the displeasure of dialectic materialism $^{[2]}$.

Another aspect of this rise in obscurantism is that, in the world of physicists, interest in the foundations of quantum mechanics has been revived, stimulated by new experiments, which make it possible to probe further into the microscopic world $^{[3]}$. However, the present generation of physicists have been brought up largely just to apply quantum mechanics and have learnt to scorn discussing the foundations thereof. The former discussions of the [nineteen] twenties had not led to satisfactorily clear statements. Bohr had indeed given correct answers to most of the questions, but his formulation was too vague to give an unambiguous procedure in all cases that occur. His attempts to incorporate `complementarity' as a general fundamental principle of human thought unnecessarily increased the confusion, including among his followers. Moreover there were opponents ( Einstein, de Broglie , Bohm ) who found the Copenhagen interpretation unbearably dogmatic and sought a more classical tangible reality behind the phenomena.
\\
Such a clash of opinions within an exact science would only be astonishing, if we had not been familiar with it from earlier instances - when Newton introduced a hidden force that acted at a distance; when in the nineteenth century atoms and molecules were postulated; and when Einstein undermined the idea of absolute time. It is not strange or inappropriate that a new theory forces one to view phenomena in a new and unexpected way. It is, however, indeed inappropriate to mention such a `change of paradigm' as if on a par with a change in politics or women's fashion. For sure, the new theory and accompanying new interpretation are an ineluctable consequence of the phenomena. At the same time, however, that means that \emph{each difference of opinion concerning the interpretation can be settled by getting to grips with the actual phenomena}.
\\
The classical equations of motion for a particle are expressed in terms of its coordinates and momenta, and these can be observed directly. On the other hand, quantum mechanics describes the particle using a wave function and one must ask oneself how observations are made. This is the problem of \emph{measurement} and is the essence of the discussion, since it is here that scientific formulation and laboratory work meet full on. My intention here is to work through the simplest of possible realistic models of the measurement process and thereby test the above three declarations on real phenomena.

\section{Model of a measuring process}

If a lightwave is incident on a screen with two slits, there appears an interference pattern on a photographic plate placed behind it. According to quantum mechanics the same goes for a beam of electrons. In other circumstances however, such as the bending of cathode rays, those same electrons display themselves as particles, each with a position $\boldsymbol{r}$ and therefore also a trajectory. Applied to the two slit experiment, this picture leads to a paradox. Each electron passes through one or the other slit; how can it be influenced by the presence of another slit? How can it be that the darkening of the plate is not the sum of those caused by the separate slits, but, instead of that, an interference pattern; and further why is the latter dependent on the distance between the slits? Bohr's answer : as long as the measurement apparatus does not allow the path to be determined, it makes no sense to talk about a path. As soon as one chooses an experimental apparatus that is indeed able to determine through which slit the electron passes, then interference can no longer occur. I shall construct an apparatus that can observe the passage of the electron and can be described entirely by the usual laws of quantum mechanics. In this way the epistemological flavour is removed from Bohr's statement. For details see $^{[5]}$.

Place in the upper slit an atom that possesses a $1S$ ground state , but which is now excited into its $2S$ state. This state is metastable because the transition to the ground state is [dipole] forbidden. However, as the observed electron passes, the  $2S$ state becomes distorted, obtaining a $P$ component, via the transient Coulomb field [ of the electron ] and deexcitation is possible via photon emission. Thereby the passage of the electron is observed and recorded. One can, if desired, trap the photon on a photographic plate or afterwards check the state of the measuring atom. The essential point is that the apparatus is left permanently in another state [ and not the original one].

To forestall any misunderstanding : if no photon is emitted, one cannot as a result conclude that the electron went through the other slit. Also, if one closes the second slit, so that the electron certainly went through the first, there is a chance that the atom has not been deexcited. The efficiency of the measuring apparatus depends on the nature of the atom, the width of the slit, the speed of the electron etc., and is never $100$ \%. That only produces inessential complications which I will not go into here. Another caveat is needed in order to reasonably state that our observation may be described as a measurement. Each measurement of a variable can be formulated as a series of questions : “Does the value of the variable lie in the interval $(0,\Delta)$ ?" ; “Does the value lie in  $(\Delta,2\Delta)$ ?" ; and so on. Conversely, each question about the state can be formally regarded as the measurement of a variable, namely the variable which by definition is $1$ for the state for which the answer is `yes' and defined as $0$ if the answer if `no'. For example, the question whether a system is in its ground state , $|\Phi\rangle$ , corresponds to the measurement of the variable that is represented by the operator  $|\Phi\rangle\langle \Phi|$, that is the projection operator for the state. Thus may the question whether the electron has passed through the upper slit indeed be regarded as an example of a measurement.

\section{The measurement}

The description of an electron that passes by an atom and eventually results in the emission of a photon, is a standard problem in scattering theory. The atom is initially in the $|2S\rangle$ state without the presence of a photon. After the electron has passed by, the atom can still be in the same state or it can be in the state $|1S;\boldsymbol{k}\rangle$ i.e. with the atom in its ground state and with a photon $\boldsymbol{k}$ present. Each state of my apparatus is a linear superposition of all these states. Also, to include the electron in the description, we must take the coefficients as functions of the electron coordinate $\boldsymbol{r}$ :

\begin{equation*}
    \Psi = \phi(\boldsymbol{r})|2S\rangle + \sum_{\boldsymbol{k}}\psi_{\boldsymbol{k}}(\boldsymbol{r})|1S;\boldsymbol{k}\rangle
\end{equation*}

This is the state vector for the whole system, that is to say the combination of the observed system ( that is the electron ) and the measuring apparatus ( that is to say the atom together with all the oscillators , $\boldsymbol{k}$ , of the electromagnetic field).

The evolution of $\Psi$ is given by the Schr{\"o}dinger equation   $i  \dot {\Psi} =H\Psi$, in which $H$ is the Hamiltonian for the total system. This is the sum of the kinetic energy of the electron , the excitation energy of the atom ( which has only two states of importance ), the energy of the field ( for which only the 0 and 1 photon states are included), and the interaction term. [Translator's note: it would appear that the interaction term is meant to refer collectively to both the interaction between the electron and the atom and that of the atom with the electromagnetic field.] This can only happen if the electron is in the neighbourhood of the atom and consequently the [part of] the interaction term [ corresponding to the interaction between electron and atom] contains a factor $u(\boldsymbol{r})$ , which only differs from zero if $\boldsymbol{r}$  is in the vicinity of the atom.

Long before the electron reaches the screen ( say at $t=-T$ ) the atom is certainly in the state  $|2S\rangle$. So $\psi_{\boldsymbol{k}}(\boldsymbol{r},-T)=0$, while $\phi(\boldsymbol{r},-T)$ is a given incident wave. Solve the Schr{\"o}dinger equation with these initial conditions. When the electron has completely passed by, ( say at $t=+T$ )  the state is described by $\phi(\boldsymbol{r},+T)$ and $\psi_{\boldsymbol{k}}(\boldsymbol{r},+T)$, which respectively correspond to the possibility that the electron leaves it [ the atom] behind in its original state $|2S\rangle$ and to the possibility that deexcitation has occurred with the emission of a photon $\boldsymbol{k}$.

The above conclusion can already be made without the complete solution of the Schr{\"o}dinger equation. The wave $\psi_{\boldsymbol{k}}(\boldsymbol{r},t)$ was originally zero, but becomes non zero due to the presence of $\phi(\boldsymbol{r},+t)$ and the interaction. Since this is restricted to the vicinity for which $u(\boldsymbol{r}) \ne 0$, $\psi_{\boldsymbol{k}}(\boldsymbol{r},t)$ is \emph{ a wave that fans out from the vicinity of the atom}, and hence from the upper slit. Such a wave produces a dark spot on the photographic plate , without interference fringes.

The purpose of this somewhat technical exposition was to make clear that \emph{the measuring process is completely described by the Schr{\"o}dinger equation of the total system }, object plus apparatus. No extra axiom, such as postulated by van Neuman $^{[6]}$, is needed. The intervention that declaration (i) talks about is restricted to the setting up of the experiment. The measurement then proceeds according to the normal laws of quantum mechanics.

We have found that the measuring process either registers that the electron has gone through the upper slit or does not do so. If a registration occurs, the result is permanently recorded, objective and immutable by whoever subsequently reads it. Declaration (ii) is untrue for observations that can actually be made. If the emission of a photon , $\boldsymbol{k}$, is registered, the subsequent state of the electron is described by $\psi_{\boldsymbol{k}}(\boldsymbol{r},t)$ which produces no interference. That has nothing to do with the knowledge of the observer : even if he does not look at the result, $\psi_{\boldsymbol{k}}(\boldsymbol{r},t)$ still takes on the role of describing the electron state, because that is merely a physical consequence of the interaction with the measuring apparatus. Declaration (iii) must thus be answered as follows. As long as one describes the observed system in isolation, the `reduction of the wavefunction' is not to be derived from the equation of motion and must be introduced as an extra postulate. But if one combines it [ the observed system] with the apparatus, the reduction is  a direct consequence of the total Schr{\"o}dinger equation for this combined system. As soon as the apparatus measures something, it evolves into a new state and a new coefficient goes with it, which becomes the new wavefunction for the electron.

\section{Macroscopic Systems}

The permanent registration came about because the atom was prepared in a metastable state and therefore could make an irreversible transition. Irreversibility appears only in macroscopic systems , systems that have very many degrees of freedom, which can be treated with the help of statistical mechanics. Our measuring apparatus can be considered as macroscopic, because the electromagnetic field has numerous degrees of freedom and the emission process is consequently irreversible. Generally : a quantum mechanical measurement takes place when a macroscopic measuring apparatus, prepared in a metastable state, is pushed out of that state by a microscopic influence: Wilson cloud chamber, Geiger counter etc.

Macroscopic systems have the property that their energy levels lie unimaginably close to each other, so that such a system never exists in a single eigenstate. Because of that they appear to us to behave very differently from what we are accustomed to expect from atoms and other small objects, for which quantum mechanics was originally developed. \emph{For sufficiently large systems, indeed, it turns out that the quantum mechanical treatment leads to a classical description using probabilities $^{[7]}$ }, so that statistical mechanics is valid for them. That makes metastable states possible and consequently the macroscopic, permanent and objective recording of events which take place on the microscopic scale. The different positions of a pointer on a voltmeter are not different eigenstates of the apparatus : each position is a collection of unimaginably many eigenstates.

The point of view deserves further comment. Characteristically, quantum mechanics, besides probabilities , also features cross terms , which contain the interference of the probability amplitudes. The italicised statement above contains the fact that for macroscopic systems the cross terms [taken all together] are zero, or at least in practice removed, just as the interference terms prevent the interference of light from two incoherent sources. In passing, this is also the solution to the riddle of Schr{\"o}dinger's cat : it is indeed  certainly  dead or alive; the paradoxical interference terms between both possibilities are negligible because the cat is macroscopic. Willful mutilation of quantum mechanics to make these terms zero $^{[8]}$ is superfluous.

Microscopic equations of motion, such as the Schr{\"o}dinger equation for $\Psi$, are invariant under time reversal, but the transition of the macroscopic measuring apparatus from its metastable initial state to its stable final state is irreversible. \emph{This is the reason that the quantum mechanical measurement is irreversible.} It goes hand in hand with the entropy increase equal to the thermodynamic entropy change between the stable final state and the metastable initial state. Whoever wants to talk about information, must take into account that this transition goes hand in hand with an enormous loss of information, because the initial state is more narrowly confined in phase space than the stable final state. This thermodynamic information loss totally overwhelms that gained from knowledge that the electron has passed through the upper slit. 

In our model the initial state of the apparatus consisted of a single microscopic state $|2S\rangle$ and the final state consists of a superposition of the states $|1S;\boldsymbol{k}\rangle$. The entropy increase comes about because no distinction is made between the different $\boldsymbol{k}$. Look now at the electron rather than at the apparatus. Its initial state is a single Hilbert vector $\phi$ ; its final state, that after the photon has been emitted, is the subspace spanned by all the $\psi_{\boldsymbol{k}}$ , because the photon is not specified. Because this final state is less specific than the initial state, the entropy is increased. Even for the apparatus the increase is caused by the mutual lack of distinction between the $\boldsymbol{k}$; this entropy increase then appears to be equal to that of the apparatus $^{[5]}$. One also now sees how it is possible that the pure initial state $\phi$ transforms into a mixture. That could not happen if the electron obeyed the Schr{\"o}dinger equation by itself, but its wavefunction is not autonomous  but a subspace of the total wavefunction $\Psi$.

\section{Final Remarks}

A measurement is a normal physical process, in which the system of interest interacts with an appropriately chosen measuring apparatus. On those grounds other questions can be investigated, besides the three declarations [i.e. those in the introduction] .

Is it true that each measurement splits the universe into many countless, equivalent, even actually existing, samples? This idea has been, in all seriousness, proposed as a way out of the difficulties with the measurement process $^{[9]}$. Putting aside the fact that this proposal is neither verifiable not falsifiable, it appears also not definable, because a measurement is not to be distinguished from the continuous welter of natural processes.

Is it true that in order to read a measuring apparatus, a second apparatus is needed, and so forth, up to the brain of the observer $^{[10]}$? No, the measurement is complete the moment the result is macroscopically recorded, so that someone afterwards can look at it without disturbing it. Where does the border between the microscopic and macroscopic lie?( A familiar sophism is to argue that the distinction does not exist.) Answer: a system is macroscopic if, within the required accuracy, its behaviour satisfies a macroscopic description. For something to serve as a measuring apparatus, it must moreover have different states that exist long enough for our purpose and stable enough so that it can be read without it being disturbed. Whoever demands a more absolute characterisation does not know what physics is.

Must a measurement be instantaneous? This is what for simplicity von Neumann had assumed, but it is not clear how his postulate of the reduction of the wavefunction can be other than instantaneous. We know now that the duration of the measurement is the time for which the interaction term is different from zero. In our model the measurement was a scattering process and the duration was therefore automatically determined. Before and afterwards the electron was free; the reduction of the wavefunction is found looking back to be the difference between the incident and outgoing waves. How the reduction gradually came into being during the measurement was described in detail by the Schr{\"o}dinger equation for $\Psi$.

Is the uncertainty principle broken by `experiments with negative results', that is measurements that demonstrate no change $^{[11]}$? The possibility of a negative result is described by the component $\phi(\boldsymbol{r},+T)$ of our outgoing wavefunction. Given that it emerged from a well defined Schr{\"o}dinger equation, one knows without further analysis, that the  uncertainty principle is safe - just as safe as for any scattering process.

Yet one more last cautionary word, about a subject that hardly sits naturally with the measurement process, but has recently received a lot of attention. In 1935 Einstein, Podolski and Rosen proposed a thought experiment, that never claimed to disprove quantum mechanics, but made very clear how much it was at odds with what classical physics expects and what Einstein demanded as a minimum requirement for a theory. (Pauli spoke of `Einstein's neurotic misunderstanding'.) With an ingenious calculation, J.S.Bell proved that the quantum mechanical result was inconsistent with any attempt to give quantum mechanics a classical background with the help of hidden variables - unless one allows that they interact instantaneously at a distance, and so with infinite speed. (That is not necessarily a violation of relativity theory because it is not possible by this means to achieve signalling.)

In 1982 the experiment was carried out by Aspect and others with the result that quantum mechanics was confirmed, which amazed no one. It is a useful experiment, carried out with great skill, but it is nonsense to label it as `one of the most significant discoveries of the century'$^{[12]}$. It is more appropriate to compare it to Foucault's pendulum experiment, which confirmed something already known, than with the experiment of Michelson and Morley, which revealed an unknown world. Also Bell's result is not `one of the most remarkable results of the twentieth century'$^{[13]}$ Unsuccessful experiments, that is to say experiments that do not give the expected results, are much more important than the successful ones which give precisely what one expected. The whole episode of the EPR paradox leads only to the conclusion that quantum mechanics is unsettling, but inescapable, unless one stubbornly clings to the hidden variables of de Broglie and Bohm, and accepts into the bargain [takes for granted] that they interact instantaneously$^{[14]}$.

\section{References}

1. M. Jammer \textit{The Philosophy of Quantum Mechanics} (Wiley-Interscience, New York 1974); L.E. Ballantine, Amer. J. Phys. $\boldsymbol{55}$, 785 (1987).\\2. L.R. Graham, \textit{Science and Philosophy in the Soviet Union} (Allen Lane, London 1973)\\3. D.M. Greenberger ed., \textit{New Techniques and Ideas in Quantum Measurement Theory} ( New York Acad. Sci., New York 1986 ).\\4. P.C.W. Davies and J.R. Brown eds. \textit{The Ghost in the Atom} (Cambridge Univ. Press 1986).\\ 5. N.G. van Kampen, Physica A $\boldsymbol{153}$, 97 (1988)\\ 6. J. von Neumann, \textit{Mathematische Grundlagen der Quantenmechanik} (Springer, Berlin 1932)\\7. N. G. van Kampen, Physica $\boldsymbol{20}$, 603 (1954)\\8. G.C.Ghirardi, A. Rimini, and T. Weber, Phys. Rev. D $\boldsymbol{34}$, 470 (1986).\\9. B.S. DeWitt and N. Graham eds., \textit{The Many-Worlds Interpretation of Quantum Mechanics} (Princeton Univ. Press, Princeton 1973).\\10. F. London et E. Bauer, \textit{La th{$\acute e$}orie de l'observation en m{$\acute e$}chanique quantique} (Hermann, Paris 1939); E.P. Wigner, \textit{Symmetries and Reflections} (Indiana Univ. Press, Bloomington, 1967) p 171.\\ 11. M. Renniger, Z. Physik $\boldsymbol{158}$, 417 (1960).\\ 12. M. Talbot, \textit{Beyond the Quantum} ( Bantham Books , Toronto 1987) p 1.\\ 13. E. Squires, \textit{ The Mystery of the Quantum World} ( Adam Hilger, Bristol 1986)\\ 14. J.S. Bell, Lecture at Groningen 16 October 1989.

\begin{center}
    \textbf{\large{End of the translation of “Het Postmoderne Obscurantisme"}}
\end{center}

\begin{center}
    \textbf{\large{Translator's Acknowledgements}}
\end{center}

I am grateful to Prof Brad Foreman for suggesting, on seeing a draft, that such a translation be published and also for many helpful discussions on the quantum measurement problem generally. I also thank Prof Ubbo Felderhof for valuable comments and suggestions on the translation in an earlier version of the manuscript. Any shortcomings in the present version, of course, are the responsibility of the translator. And I thank Prof Stewart Clark for his interest and support.\\
This translation is published with the permission of the executor,
Mr drs. A.J.van der Sloot, of Prof N G van Kampen's estate whose ready willingness to help is much appreciated. I am also grateful to Prof Gerard 't Hooft, as nephew to Prof N G van Kampen, for correspondence on his aspirations for the proposed translation.

\end{document}